# Detection of Interlayer Interaction in Few-layer Graphene through Landau Level Spectroscopy

Zefei Wu[†,⊥], Yu Han[†,⊥], Wei Zhu[‡], Mingquan He[†], Shuigang Xu[†], Xiaolong Chen[†], Weiguang Ye[†], Tianyi Han[†], Huanhuan Lu[†], Rui Huang[§], Lin Wang[∥], Yuheng He[†], Yuan Cai[†], Rolf Lortz[†] and Ning Wang[†,*]

[†]Department of Physics and the William Mong Institute of Nano Science and Technology, the Hong Kong University of Science and Technology, Hong Kong, China

[‡]Department of Physics and Astronomy, California State University, Northridge, California 91330, USA

[§]Department of Physics and Electronic Engineering, Hanshan Normal University, Chaozhou, China

[∥]Department of Condensed Matter Physics, Group of Applied Physics, University of Geneva, Switzerland

ABSTRACT

We demonstrate that surface relaxation, which is insignificant in trilayer graphene, starts to manifest in Bernal-stacked tetralayer graphene. Bernal-stacked few-layer graphene has been investigated by analyzing its Landau level spectra through quantum capacitance measurements. We find that in trilayer graphene, the interlayer interaction parameters were similar to that of graphite. However, in tetralayer graphene, the hopping parameters between the bulk and surface bilayers are quite different. This shows a direct evidence for the surface relaxation phenomena. In spite of the fact that the Van der Waals interaction between the carbon layers is thought to be insignificant, we suggest that the interlayer interaction is an important factor in explaining the observed results and the symmetry-breaking effects in graphene sublattice are not negligible.

TEXT

Surface relaxation (contraction of the interlayer spacing of the surface layers) is widely present in bulk crystals.[1-3] In bilayer graphene (the thinnest system with interlayer coupling), an expansion of the layer spacing was observed in comparison with graphite.[4,5] In trilayer graphene, based on the study of Landau level spectrum, the interlayer coupling that is characterized by the Slonczewki-Weiss-McClure parameters is consistent with those of bulk graphite, indicating that the interlayer spacing neither expands nor contracts.[6-8] The question whether surface relaxation exists in few-layer graphene with interlayer coupling via the Van der Waals interactions is well worth studying because a tiny relaxation may induce fluctuation of the interactions and thus result in a pronounced modulation of Landau level characteristics. Few-layer graphene consists of stacked single layer graphene sheets through the Van der Waals (VdW) interaction. Although the interaction between graphene layers is weak, two stacking orders are energetically the most-favorable states, namely the Bernal stacking (ABA stacking, the most common form in nature) and the rhombohedral stacking (ABC stacking, a relatively rare form).[9-15] The finite number of stacked graphene layers provides another degree of freedom to tune the properties of graphene meanwhile the two-dimensional (2D) nature of graphene still remains.[6-8,11,14,16-25] Compared to single-layer graphene, the electronic band structure of few-layer graphene is complicated because different stacking orders would give rise to certain types of chirality for Dirac Fermions. Even for the simple Bernal stacking, the interlayer interactions can lead to emergence of new Dirac points and band-shiftingin few-layer graphene samples.[25-27] Moreover, many-body effects, such as electron-electron interactions that have been reported in single-layer graphene[28,29] near the charge neutrality point, may play a significant role in few-layer graphene.

Few layer graphene with Bernal stacking order, recent studies based on the single-particle picture have showed that the interactions between graphene layers are weak but essential in determining the band structures in the low energy regime.[30] The interlayer interactions can be parameterized as higher-order hopping energies in the tight-bind (TB) model known as the Slonczewki-Weiss-McClure (SWMcC) parameters that have been previously defined in graphite.[31] Recently, it has been demonstrated that these parameters in trilayer graphene are slightly different from those in graphite, indicating that the surface layers should have a considerable influence on the properties of few-layer graphene systems.

Apparently, for one- or two-layer graphene samples, there is no bulk component. However, for three or more layers (before reaching the bulk limit), the interlayer separation in surface layers may be slightly smaller than that in the bulk because of distinct interlayer interactions in the surface few layers. Since these interactions lift the degeneracy of zero-energy Landau level (LL) of Bernal-stacked few-layer graphene, even slight corrections can give rise to pronounced changes in quantum phase transition in the low energy regime. Therefore, it is expected that the LL spectrum should be modulated due to the surface relaxation[2,4,5,32] in few-layer graphene. In this work, we focus on trilayer and tetralayer graphene with Bernal stacking order, the layer numbers at which the bulk component emerges. By comparing the SWMcC parameters extracted from quantum capacitance (QC) spectra, we found that in tetralayer graphene the interlayer interaction between surface and bulk layers is larger than that between two neighboring layers in bulk graphite. However, in trilayer graphene, the surface relaxation turns out to be insignificant. The present work may shed light on the study of other two dimensional materials as well.[33,34]

Trilayer and tetralayer graphene samples were exfoliated onto $Si/SiO_2$ substrates for optical identification and further Raman spectroscopy studies.[35] The profiles of Raman 2D peaks can be used to distinguish Bernal and rhombohedral stacking orders. Typically, the 2D peaks of Bernal-stacked trilayer and tetralayer graphene are more symmetric than that of rhombohedral-stacked graphene.[36] To prepare graphene capacitors, we used atomically thin flakes of hexagonal boron nitride (hBN) as the dielectric materials.[37] After aligning the hBN flakes onto graphene by the dry-transfer method (described in the Supplementary Information and Methods), top-gate electrodes are prepared on the hBN flakes.[38] Different from the quantum Hall effect (QHE) which reveals the chiral edge states of graphene, the QC method yields the averaged density of states in graphene devices.[39] Moreover, the measurement of QC spectra provides a more convenient and straightforward technique to spatially distinguish few-layer graphene, characterize the number of graphene layers, stacking order as well as the interlayer interactions.

Fig. 2a and Fig. 3a show the measured total capacitance $C_{tot}$ of trilayer and tetralayer graphene as a function of top-gate voltage $V_{tg}$ respectively. The oscillation of $C_{tot}$ is due to highly degenerated LLs. The minima in the oscillation are corresponding to those filled LLs with filling factors described by $\nu = nh/eB$. In the region of high chemical potential, the filling factor increases by a factor of four because of spin and valley degeneracy. Therefore, the spacing between adjacent minima $\Delta V_{tg}$ can be used to extract the capacitance of hBN by $C_{BN}(\Delta V_{tg} - \Delta E_F/e) \approx C_{BN}\Delta V_{tg} = 4Se^2B/h$. The high-quality thin hBN flakes enable us to apply gate voltages up to $\pm 5\ V$ without breaking down their dielectric property and thus observe LLs with large indices. We found that some LLs cross at certain magnetic fields (B-fields), very different from that in signal layer graphene (SLG) and bilayer graphene (BLG) samples. This unique feature should originate from the hybridization and intersection of LLs developed from SLG-like subband or BLG-like subband in trilayer and tetralayer graphene. Another interesting feature is the symmetry breaking states at zero energy LL which is attributed to the chirality of Dirac Fermions in Bernal-stacked few-layer graphene (l= 1, or l = 2, or both).

To further characterize the fine band structures and LL fan diagrams of trilayer and tetralayer graphene, calculations based on the TB model have been carried out by adjusting the SWMcC parameters. The interactions between graphene layers can be numerically parameterized as six hopping energies $\gamma_0, \gamma_1, \gamma_2, \gamma_3, \gamma_4$ and $\gamma_5$ as shown in Figure 1e. The on-site energy difference $\delta$ (the difference between two honeycomb sublattices in the same layer) is also accounted for completeness. The nearest intralayer hopping $\gamma_0$ is fixed (3.1 eV) as that in SLG while the hoppings between nearest and next-nearest layers are variables in order to fit the experimental data. For trilayer graphene, the top and bottom layers are two surfaces with respect to the middle one. The interactions between the neighbor layers $\gamma_1, \gamma_3$ and $\gamma_4$ are set to be the same both for top-to-middle bilayer and for middle-to-bottom bilayer. For tetralayer graphene, we can consider that there is a bilayer structure occurring between the top and bottom surface layers. In this case, the interactions between the bilayers (sandwiched between the top and bottom layers) should be different from that at the outermost surface layers. Therefore, we label the interlayer interactions $\gamma_1, \gamma_3, \gamma_4$ for outermost surface bilayers and $\Gamma_1, \Gamma_3, \Gamma_4$ for the innermost ("bulk") bilayer. Parameters $\gamma_2$ and $\gamma_5$ are for interactions between first and third layer which are essential for symmetry breaking states. The definition of these parameters is illustrated in Figures 1e and 1f. First, we use the parameters obtained from graphite to perform numerical calculations for the LL spectrum. The density of states (DOS) ($\Delta$ is LL broadening) is described by: $\rho(E; E_{LL}) = \frac{2eB}{h} \frac{1}{\pi} \frac{\Delta/2}{(E-E_{LL})^2+(\Delta/2)^2}$. The QC of graphene is proportional to DOS ($C_q = e^2 \rho(E)$). The total measured capacitance $C_{tot}$ can be obtained by series combination of the geometrical capacitance $C_{BN}$ and quantum capacitance $C_q$, $C_{tot} = (C_q^{-1} + C_{BN}^{-1})^{-1}$. In order to compare simulation results with experimental data, we convert the chemical potential E to $V_{tg}$ based on the charge conservation in graphene capacitors: $C_{BN}\left(V_{tg} - \frac{E_F}{e}\right) = e\int_0^{E_F} \rho(E)dE$ .[40,41] Consequently, we established a direct link between a given set of SWMcC parameters to the relationship of $C_{tot}$ vs $V_{tg}$. By this way, we deduce the parameters by varying their values around those of graphite so that the simulated $C_{tot}$ vs $V_{tg}$ can be best fitted with experimental results.

For trilayer graphene, the interlayer interaction parameters are fitted using LL crossing positions. The best fitting yields $\gamma_1 = 0.37$ eV, $\gamma_2 = -0.032$ eV, $\gamma_3 = 0.30$ eV, $\gamma_4 = 0.04$ eV, $\gamma_5 = 0.05\ eV$ and $\delta = 0.04$ eV. It should be noted that $\gamma_3$ and $\gamma_4$ are close to that of graphite because they are less sensitive in determining LL crossings than the rest parameters. In the low energy region, the energies of nearly disperseless bands at the LL with index $N = -1, 0$ are determined by $\gamma_2, \gamma_5$ and $\delta$ collectively, which effectively split the zero-energy LL. In the high energy region, SLG-like bands with dispersion E~$\sqrt{B}$ cross with BLG-like bands with dispersion E~$B$ (see figure 2c). As $\gamma_0$ and $\gamma_1$ are responsible for the shapes of each subband and $\gamma_2, \gamma_5$ and $\delta$ determine the subband offset, therefore, all the parameters have to be taken into account simultaneously in order to locate the positions of these crossing points. Experimentally, the crossing positions are observed at $V_{tg} \approx 0$, $V_{tg} > 3\ V$ and $V_{tg} < -2\ V$. At $V_{tg} \approx 0$, an additional peak (labeled by Arrow 1 in figure 2c) emerges at $B > 7T$ and forms a disperseless band, and this peak also shows up at $B < 5T$ (labeled by Arrow 4 in figure 2c). At $V_{tg} > 3\ V$, some broadened LLs (labeled by Arrow 5 in figure 2c) are observed. The LL indices of the broadened bands decrease by increasing the magnetic field strength. It clearly implies that the $\sqrt{B}$ dependent band intersects with a set of B dependent bands. Similarly, at $V_{tg} < -2\ V$, as labeled by Arrow 3 and Arrow 4 in figure 2c, the crossing occurs but with fewer crossing points. The optimal fitting is achieved when all these crossing features can be reproduced in the simulation. The splitting of low energy LLs into two-fold degeneracy is not observed probably due to the presence of disorder and thermal excitation. In addition, the dispersionless band appearing at B = 8T can be further confirmed by the emergence of the quantum Hall plateau at a filling factor $\nu = +6$. The plateau becomes well resolved as the B-field goes off the crossing point (see Supplementary Information).

For tetralayer graphene, the interlayer interaction parameters for the outermost surface bilayers are determined to be $\gamma_1 = 0.45$ eV, $\gamma_3 = 0.35$ eV, $\gamma_4 = 0.045$ eV. While the parameters for the "bulk" bilayer are $\Gamma_1 = 0.23$ eV, $\Gamma_3 = 0.17$ eV, $\Gamma_4 = 0.023$ eV. The parameters of next nearest interlayer interactions are $\gamma_2 = -0.11$ eV, $\gamma_5 = 0.02$ eV, and $\delta = 0.025$ eV. To better understand how these parameters influence the tetralayer graphene (as listed in Table. 1), it is essential to first check some key features in the LL spectrum. In the low energy region, the 16-fold zero-energy LL strengthens and forms four four-fold bands that can be classified into two groups separated by a gap located at zero energy. The small gap (gap I and II, as shown in the inset in Fig. 3c) between the two bands in each group is too small to be clearly separated. Thus two eight-fold bands with a considerable gap can be recognized and this is confirmed convincingly from the QHE data measured from the ABAB tetralayer Hall device (see Supplementary Information) in which the transverse conductivity forms a plateau at zero filling factor. The nearest two neighbor plateaus appear at $\sigma_{xy} = -8e^2/h$ and $+8e^2/h$, respectively. The absence of $\sigma_{xy} = \pm 4e^2/h$ also agrees with the smeared small gaps at $\nu = \pm 4$. The central gap (gap III, about 10meV as shown in the inset in Fig. 3c) is determined mainly by $\gamma_2$. In our QHE measurement, we also estimated the gap through the temperature dependence of resistance at B=8 T. The gap is estimated to be about 14meV which agrees well with the calculation here. At the hole side, the fifth band is broader than the others as marked by the arrows in Fig. 3a. This is due to the overlapping of two nearly parallel bands developed individually from two BLG-like subbands. This feature cannot be explained if we assume that the interlayer interactions for outermost surface bilayer and 'bulk' bilayers are identical. For tetralayer graphene, we believe that the interlayer coupling varies with the positions of graphene layers. As illustrate in the inset in Fig. 3b, the spacing of the surface bilayer should contract to some extent while the "bulk" bilayer may expand or keep unchanged. In contrast, trilayer graphene contains only one graphene sheet in the "bulk", the interlayer interactions of the outermost surface bilayers are roughly the same as pure bilayer one.

Apart from detecting interlayer interaction in few-layer graphene, the QC spectroscopy technique is also capable of detecting the band structure changes of graphene which directly arise from the modulation of the DOS (not easily extracted from electrical transport data[37,42]) such as through resonant states and negative compressibility.[29,43-45 29,42-44 29,43-45 29,43-45] In trilayer graphene, the LL at $\nu = 0$ (the one corresponding to the band at $V_{tg} \approx -0.5\ V$) is much wider than that expected and thus shifts all the bands at the hole side negatively. This special feature (at $\nu = 0$) also manifests in the QHE data where insulating behavior is observed. This behavior is unexpected in the single-particle picture of electrons. Electron-electron interactions involving many-body effects at low carrier density may play an important role for the insulating behavior.[20,46]

In the QC spectrum of tetralayer graphene, the missing LL crossing around 6T at electron side indicates the changes of interactions between graphene layers. It possibly attributes to two factors: the effect of substrate and the short screening length of graphene. Both factors were already illustrated in multilayer graphene and MoS2 devices. $SiO_2$ substrates have been proved to have much more impairments than hBN flakes over graphene. Consequently, it is conjectured that the mobility of electrons in the bottom bilayer may be lowered down by the $SiO_2$ substrate attached[47] while the top bilayer is far less influenced because of weaker interactions with the bottom bilayer.

Meanwhile, the short screening length means the electrons populated through top gate are distributed more on the top bilayer, as gate voltage is tuned away from CNP. Thus considering the two factors, the top bilayer contributes greater to oscillatory QC spectrum than the bottom bilayer. As a result, the tetralayer graphene behaves as if a bilayer in the electron side. In the hole side, however, as the Fermi velocity of holes is in general smaller than that of electrons, the amplitude of QC oscillation in the hole side is always small, as shown in the QC spectrum in Figure 3. Moreover, we believe the top bilayer and bottom bilayer contribute roughly equally to the QC spectrum. It seems that the effects from the $SiO_2$ substrate and screening are less important in the hole side so that the LL crossing is still observable.

In summary, Bernal-stacked few-layer graphene has been investigated by analyzing its Landau level spectra through QC measurements. By fitting the interlayer interaction parameters with the SWMcC model, we observed that the interlayer interaction is important and symmetry-breaking effects in graphene sublattice are not negligible, in spite of the fact that the Van der Waals interaction between the layers was thought to be insignificant. For trilayer graphene, the interlayer interaction parameters were similar to that of graphite. However, the hopping parameters between the bulk and surface bilayers are quite different, which showed a direct evidence for the surface relaxation phenomena.

FIGURES

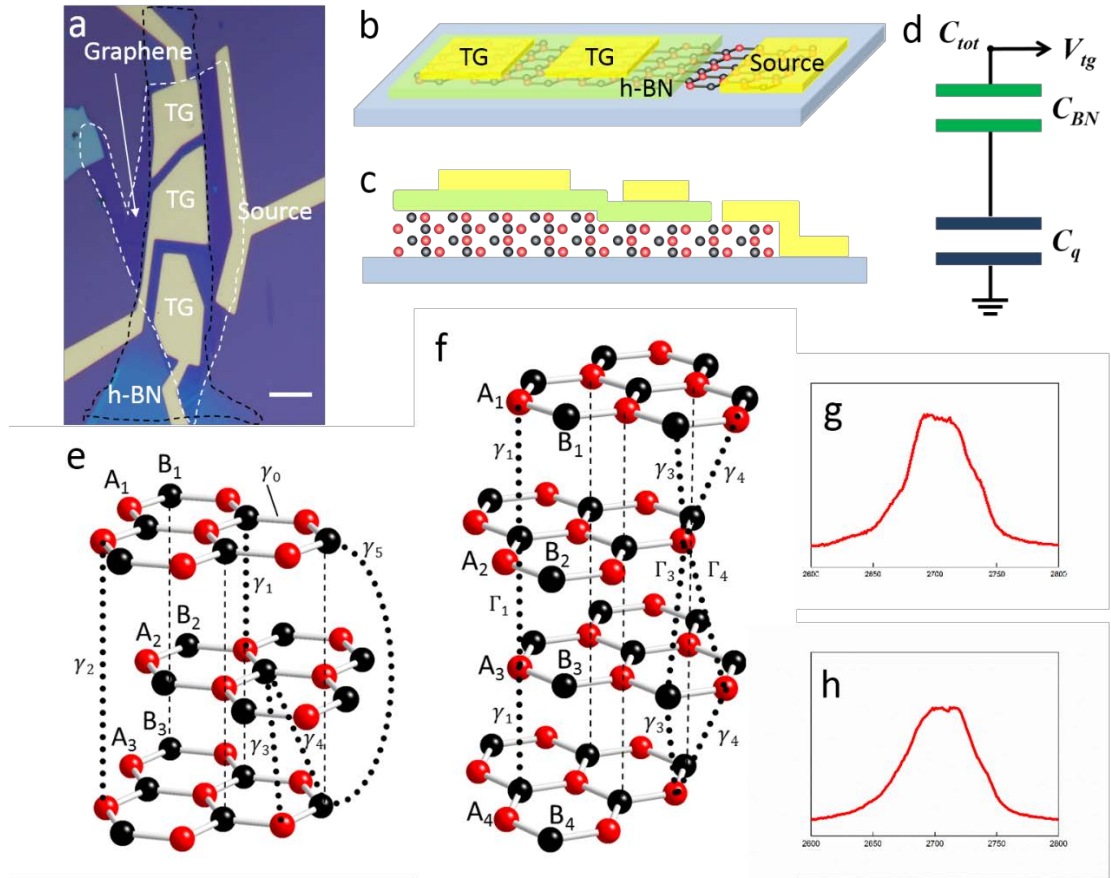

**Figure 1.** Few-layer graphene quantum capacitors. a, An optical image of trilayer and tetralayer graphene quantum capacitance devices. Our fabrication process consists of mechanically exfoliating natural graphite and synthesized hBN on SiO2/Si substrate and PMMA films separately and a dry transfer step (see Supplementary Information) to align them together using an optical microscope. The device is then annealed in Ar/$H_2$ atmosphere to remove residues on graphene and followed by standard e-beam lithography (Raith e_Line). b and c, Schematic configuration of the device. d, The equivalent circuit for capacitance measurement. e and f, Lattice structures of Bernal-stacked trilayer and tetralayer graphene. The dashed lines describe the interlayer hopping parameters. g and h, The Raman 2D bands of Bernal-stacked trilayer (g) and tetralayer (h) graphene.

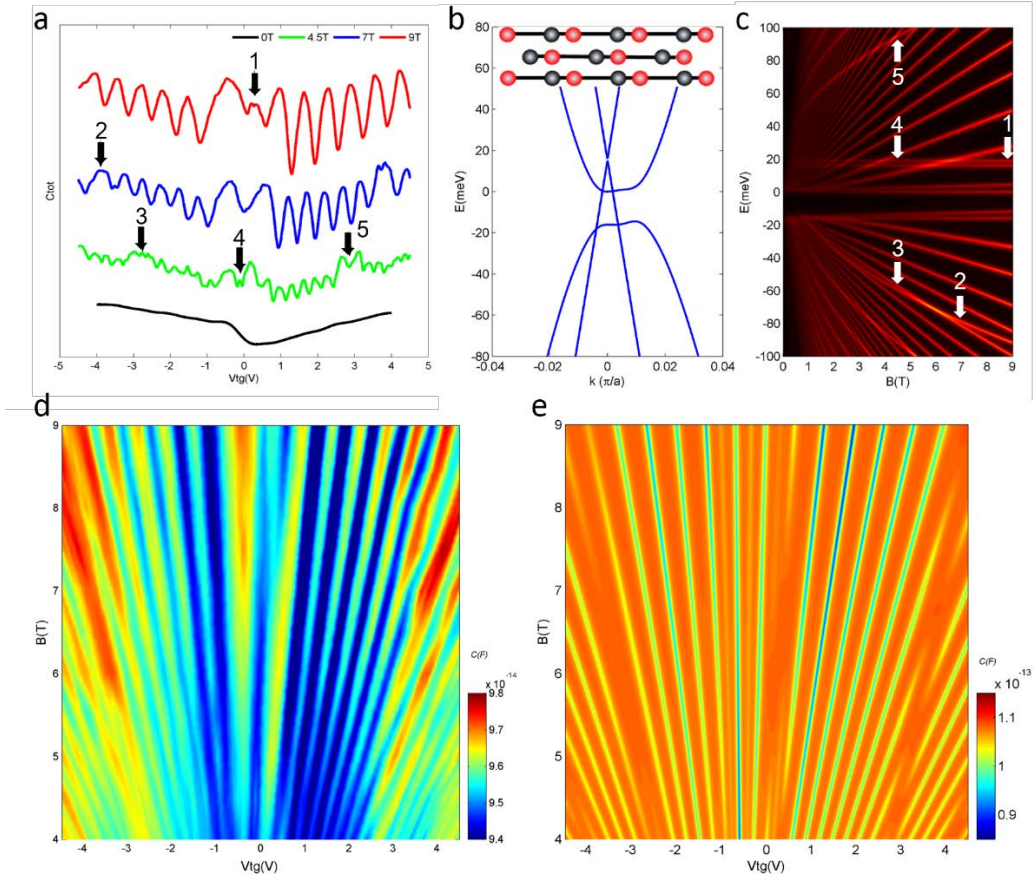

**Figure 2.** Landau level spectrum of Bernal-stacked trilayer graphene. a, Quantum capacitance of trilayer graphene as a function of gate voltages at 2K. The arrows signify the LL crossing points that evolve with B-field. b, The band structure in the vicinity of K (K') at the low energy regime. The inset shows the side view of the relaxed lattice. c, The LL spectrum of trilayer graphene as a function of B-field and energy. Two sets of LLs originated from SLG and BLG-like bands cross at certain B-fields. Note that the lowest six bands are two-fold spin-degenerated and the rest are spin- and valley-degenerated. d, Color map of measured total quantum capacitance as a function of gate voltages and B-field, the beating patterns at higher voltages and bifurcation near zero voltage imply LL crossings. e, Color map of simulated total quantum capacitance calculated based on the LL spectrum shown in c with fitted SWMcC parameters. The gate voltage is negatively shifted considering unintentional doping effects.

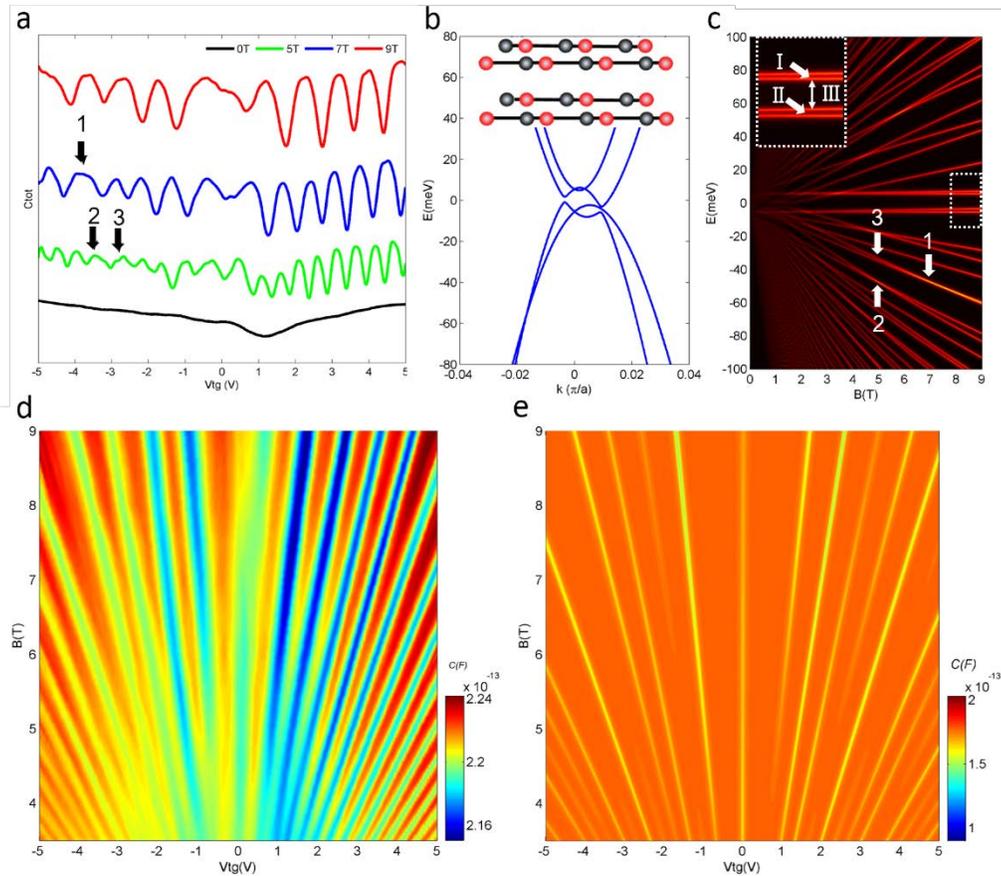

**Figure 3.** Landau level spectrum of Bernal-stacked tetralayer graphene. a, Quantum capacitance of tetralayer graphene as a function of gate voltages at a fixed B-field. The arrows signify possible LL crossing. Note that the larger peak amplitudes in electron side than hole side indicates larger Fermi velocity of electrons than that of holes. b, The band structure in the vicinity of K (K') at the low energy regime. The inset shows the side view of the relaxed lattice. c, The LL spectrum of tetralayer graphene as a function of B-field and energy. Note that the 16-fold zero-energy LL splits into four four-fold bands. d, Color map of measured total quantum capacitance as a function of gate voltages and B-field. e, Color map of the simulated total quantum capacitance.

Table 1. SWMcC parameters of few-layer graphene and graphite.

| SWMcC parameters (eV) | | $\gamma_0$ | $\gamma_1$ | $\gamma_3$ | $\gamma_4$ | $\gamma_2$ | $\gamma_5$ | $\delta$ |
|---|---|---|---|---|---|---|---|---|
| Trilayer | Surface bilayer | 3.10 | 0.37 | 0.300 | 0.040 | -0.032 | 0.05 | 0.040 |
| Tetralayer | Surface bilayer | 3.10 | 0.45 | 0.350 | 0.045 | -0.011 | 0.020 | 0.025 |
| | Bulk bilayer | 3.10 | 0.23 | 0.170 | 0.023 | | | |
| Graphite | | 3.16 | 0.39 | 0.314 | 0.044 | -0.020 | 0.038 | 0.037 |

## ASSOCIATED CONTENT

**Supporting Information**. Experimental section, transport data of trilayer and tetralayer graphene samples, and the method to determine the SWMcC parameters were included. This material is available free of charge via the Internet at http://pubs.acs.org.

## AUTHOR INFORMATION

**Corresponding Author**

*E-mail: phwang@ust.hk.

**Author Contributions**

Z. W. and Y. H. are the main contributor who initiated and conducted all the experiments including sample fabrication, data collection, analyses and band structure simulations. N. W. is the principle investigator and coordinator of this project. W. Z. provided the physical interpretation and contributed to part of the theoretical work. Other authors provided technical assistance in sample preparation, data collection/analyses and experimental setup. $^{\perp}$These authors contributed equally.

**Notes**

The authors declare no competing financial interest.

## ACKNOWLEDGMENT

We first thank Prof. Marvin L. Cohen for his constructive advice. We also thank Dr. L. Jing for related discussion and Prof. D. N. Sheng for the support on related studies. Financial supports from the Research Grants Council of Hong Kong (Project Nos. 604112, N_HKUST613/12 and HKUST9/CRF/08, HKUST-SRFI) and technical support of the Raith-HKUST Nanotechnology

Laboratory for the electron-beam lithography facility (Project No. SEG HKUST08) are hereby acknowledged

# Supplementary Information

# for

# Detection of Interlayer Interaction in Few-layer Graphene through Landau Level Spectroscopy

Zefei Wu[†,⊥], Yu Han[†,⊥], Wei Zhu[‡], Mingquan He[†], Shuigang Xu[†], Xiaolong Chen[†], Weiguang Ye[†], Tianyi Han[†], Huanhuan Lu[†], Rui Huang[§], Lin Wang[∥], Yuheng He[†], Yuan Cai[†], Rolf Lortz[†] and Ning Wang[†,*]

[†]Department of Physics and the William Mong Institute of Nano Science and Technology, the Hong Kong University of Science and Technology, Hong Kong, China

[‡]Department of Physics and Astronomy, California State University, Northridge, California 91330, USA

[§]Department of Physics and Electronic Engineering, Hanshan Normal University, Chaozhou, China

[∥]Department of Condensed Matter Physics, Group of Applied Physics, University of Geneva, Switzerland

**Sample preparation and characterization.** Graphene trilayer and tetralayer samples were mechanically exfoliated onto Si/SiO$_2$ substrates from natural graphite. The sample thicknesses were verified by an optical microscope and their stacking orders were characterized by Raman spectroscopy. The profiles of the 2D peaks in Raman data were used to distinguish Bernal stacking from rhombohedral stacking orders. Typically, the 2D peaks of Bernal-stacked trilayer and tetralayer graphene are more symmetric than those of rhombohedral-stacked ones. In our study, we choose pure Bernal-stacked trilayer and tetralayer for investigations.

**Device fabrication.** Synthetic hexagonal boron nitride (hBN) crystals were used to fabricate the insulating layers for graphene capacitance devices. Atomically thin hBN flakes have been proved to be ideal dielectric material for QC measurement. The thicknesses of hBN were around 5 to 20 nm. The alignment of hBN flakes onto the graphene samples was carried out using a dry-transfer method. (See below.) We undertake sample annealing in H$_2$/Ar to remove polymer residues. Usually, small bubbles in graphene/BN surfaces aggregated into a big one during the annealing process. We choose the regions with no bubbles for making top-gate electrodes. The metal electrodes were prepared by standard electron-beam lithography (Raith e_LiNE) and electron-beam thermal evaporation.

**Dry transfer process.** Figure S1 shows a schematic diagram of our dry transfer process. First, PVA (200nm) and then PMMA (500nm) are spin-coated onto a glass slide followed by soft baking at 80 °C for 10 min. Mechanical exfoliation of hBN on PMMA films is applied using Scotch tapes. Graphene is exfoliated on SiO$_2$ substrates and then examined by Raman spectroscopy and atomic force microscopy (AFM). The hBN flakes are identified optically and transferred onto a copper frame together with the supporting PMMA film with the help of Scotch tapes. After aligning hBN and graphene together using a microscope, we gradually lift the SiO$_2$ substrate to make graphene surface adhere to the target hBN flake. A good contact is normally achieved when a sudden change of the color of the film occurs as the air gap between the film and the substrate vanishes. The last step is critical. We gradually lower the substrate to separate the substrate from the PMMA film. Due to the Van der Waals interaction, the hBN layer stays on graphene. This dry transfer method guarantees the cleanness of the interface between hBN and graphene. The device is then annealed in Ar/H$_2$ (1:1) in order to remove PMMA residues for at least 6 hours. The electrodes finally prepared by standard e-beam lithography (Raith e_Line system).

**Capacitance measurements.** Capacitance measurements were carried out by HP 4284A LCR Meter (sensitivity ~0.1 fF). During measurement, all wires were shielded and the substrates were rounded to minimize residual capacitance. The residual capacitance in our measurement setup was at the order of 1 fF. Unlike transport measurement in QHE which detects chiral edge states of graphene, QC yields averaged density of states by employing the top-gate geometry. Therefore, it can spatially distinguish few-layer graphene by capacitance spectroscopy. Compared with a Hall bar device, the QC device is more convenient and straightforward to characterize the number of layers, stacking orders as well as surface relaxation.

**Determining the SWMcC parameters for trilayer graphene.** The parameters are varied in the vicinity of those from graphite. If we directly take graphitic SWMcC parameters to calculate the LL fan diagram, then the filling factor $\nu = +2, +4, +6, +10$ should appear when field is larger than the 7T where the LL crossing occurs, as shown in Figure S2a. However, when the field goes up to 9T, only one band comes up as shown by arrows in the center of Figure 2a, which indicates that the two two-fold bands between filling factor $\nu = +2, +10$ should be closely aligned. Therefore, the filling factors in the calculated diagram are $\nu = +2, +6, +10$. It was further confirmed by QHE (Figure S2c, S2d) that the plateau at $+6e^2/h$ which is absent below 8T become well resolved at 9T. It also indicates that QC is somehow more sensitive in detecting the details in the vicinity of LL crossing because in 8T the emergent band is already well resolved in QC spectrum. A reliable fitting result is shown in Figure S2b.

**Determining the SWMcC parameters for tetralayer graphene.** The determination of the SWMcC parameters in tetralayer graphene is more tricky because of more parameters than those of trilayer one. But similar to the case of trilayer one, $\gamma_3$ and $\gamma_4$ are less sensitive, so we first varied $\gamma_1, \gamma_2, \gamma_5$, $\delta$ and $\Gamma_1$. It is easier to start with $\gamma_1 = \Gamma_1$ and inspect how crossing points moves as $\gamma_2, \gamma_5$, $\delta$ changes individually. It is found that $\gamma_2$ is most important among these three parameters and easily to be determined because it opens a gap in the zero energy and is also responsible for the first crossing point in the electron side. The smaller value of $\gamma_2$ is expected because it agrees with transport measurement in QHE (figure S3a) where the central gap is estimated from $R_{xx} \propto e^{-E_g/2k_BT}$. The larger $\gamma_1$ and smaller $\Gamma_1$ ensures the expected positions of LL crossing at the hole side meanwhile $\gamma_2, \gamma_5$, $\delta$ need to be tuned accordingly to achieve a better replicate of LL crossing. After all the other parameters are fixed, $\gamma_3, \gamma_4$ and $\Gamma_3, \Gamma_4$ are tuned proportionally to the ratio of $\gamma_1$ and $\Gamma_1$ to the reference values of graphite. Finally, all the parameters are tuned simultaneously in the vicinity of the pre-determined values in order to find an optimal combination.

**Setp 1** PMMA film is spin coated on to glass slide.

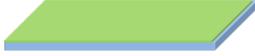

**Step 2** BN flakes are exfoliated on PMMA by scotch tape method.

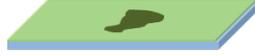

**Step 3** PMMA film with selected BN flake is transferred onto copper frame.

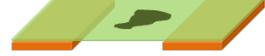

**Step 4** Copper frame is turned upside down. BN is aligned on top of graphene.

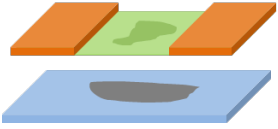

**Step 5** Substrate containing graphene is lifted and get contact with the BN flake.

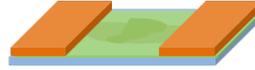

**Step 6** Substrate is gradually lowered and BN flake is left on top of graphene.

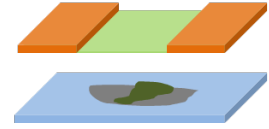

**Figure S1**. The schematic diagram for the transfer method used in the present work.

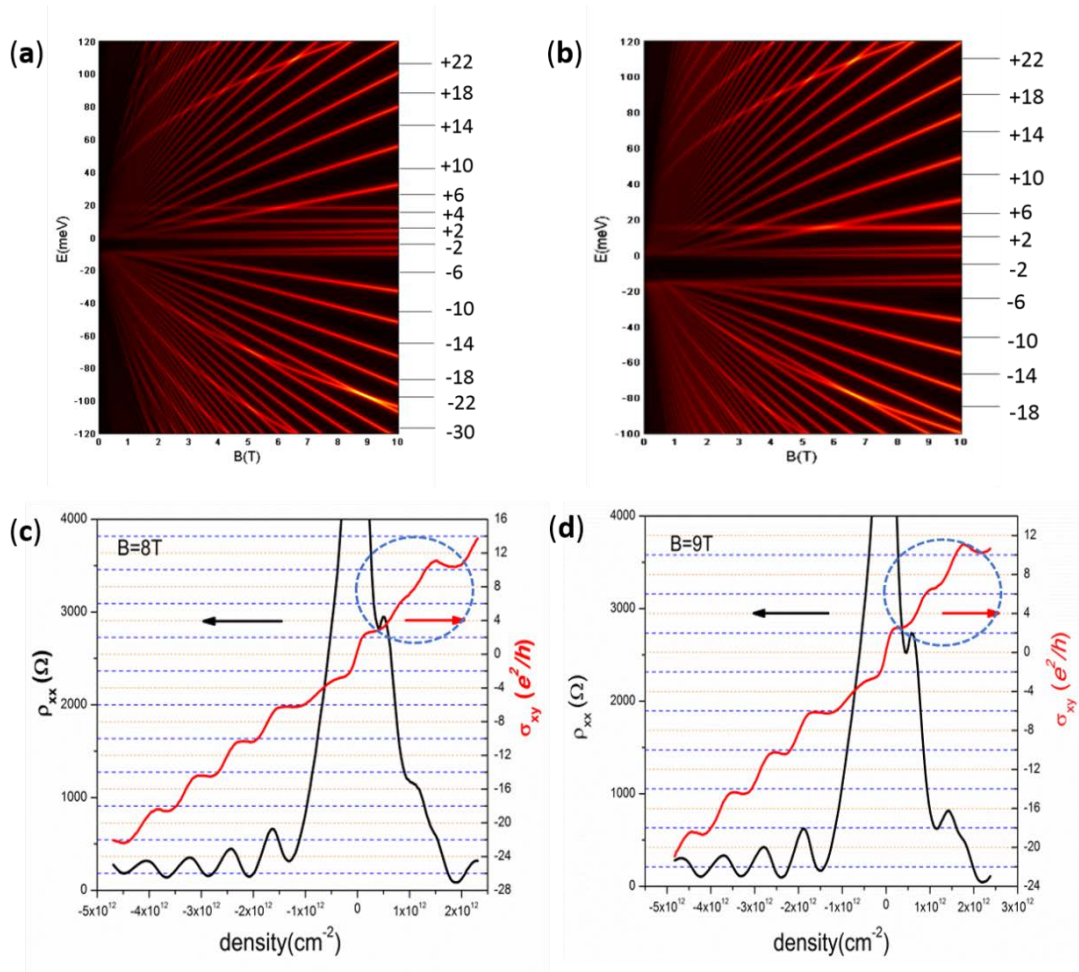

**Figure S2.** (a), (b) The LL spectra calculated using the SWMcC parameters from graphite and with fitted parameters respectively. The integers indicate the filling factors. (c), (d) QHE data obtained at 8T and 9T respectively. The absence of $+6e^2/h$ at 8T signifies LL crossing.

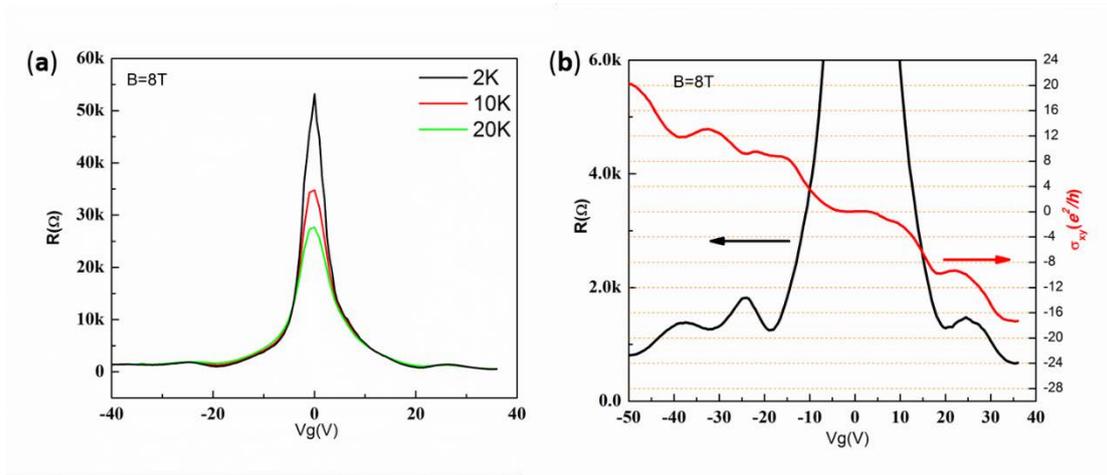

**Figure S3.** (a) The temperature dependence of Rxx at B=8T yields $\gamma_2 = 14\text{meV}$. (b) QHE data obtained at B=8T. The plateaus at low energy regions are roughly located at 0, and $\pm 8e^2/h$.